# Anisotropic crystallization in solution processed chalcogenide thin film by linearly polarized laser

*Tingyi Gu[1,2], Hyuncheol Jeong[3], Kengran Yang[4,5], Fan Wu[1], Nan Yao[1],*

*Rodney D. Priestley[1,3], Claire E. White[4,5] and Craig B. Arnold[1\*]*

[1] Princeton Institute for the Science and Technology of Materials, Princeton University, Princeton, New Jersey, 08544

[2] Electrical and Computer Engineering, University of Delaware, Newark, DE 19716

[3] Chemical and Biological Engineering, Princeton University, Princeton, New Jersey 08544

[4] Civil and Environmental Engineering, Princeton University, Princeton, New Jersey 08544

[5] Andlinger Center for Energy and the Environment, Princeton University, New Jersey 08544

*Email: cbarnold@princeton.edu

**Abstract:** The low activation energy associated with amorphous chalcogenide structures offers broad tunability of material properties with laser-based or thermal processing. In this paper, we study near-bandgap laser induced anisotropic crystallization in solution processed arsenic sulfide. The modified electronic bandtail states associated with laser irritation lead to a distinctive photoluminescence spectrum, compared to thermally annealed amorphous glass. Laser crystalized materials exhibit a periodic subwavelength ripples structure in transmission electron microscopy experiments and show polarization dependent photoluminescence. Analysis of the local atomic structure of these materials using laboratory-based X-ray pair distribution function analysis indicates that laser irradiation causes a slight rearrangement at the atomic length scale, with a small percentage of S-S homopolar bonds converting to As-S heteropolar bonds. These results highlight fundamental differences between laser and thermal processing in this important class of materials.

**Introduction:**

Chalcogenide materials have the ability to transform among various states, ranging from random amorphous networks to atomically ordered crystalline states when exposed to modest energy such as that associated with optical or thermal processes [1-4]. These different structures typically exhibit significant optical refractive index contrast, as well as differences in electronic conductivities and other material properties. Among various tuning schemes, laser processing is especially interesting as it enables precise control of the amount of energy delivered into a specific targeted region. Such methods have been explored for tuning integrated photonic devices [5-12] as well as for controlling dichroism or polarization dependent photodarkening phenomena in these materials [13-17]. It has been shown that the precise atomic structure depends on the laser wavelength, duration and intensity of laser irritations [17].

In the case of solution processed chalcogenide materials [4, 11], laser heating from above band-gap illumination has been shown to remove excess solvent related components to densify films and modify the surface chemistry and surface morphology [3]. It has been shown in arsenic sulfide ($As_2S_3$) that laser heating can induce thermal effects capable of breaking the chalcogenide-solvent bond. Furthermore, laser exposure can affect the structural properties of chalcogenide materials through polarization dependent athermal phenomena [10]. Such process can cause redistribution of atomic clusters or chemical reactions leading to anisotropic properties [13-17].

In this paper we show near-band gap laser irradiation can lead to structural changes unlike those associated with thermal processing. Beginning with a solution processed glass $As_2S_3$ film, laser processing results in crystallization and corresponding anisotropies in optical properties. Photoluminescence (PL) measurements show well-defined shifts in emission peaks relative to thermally annealed materials while PDF analysis highlights the changes in bonding structure within the films.

$As_2S_3$ solution is prepared as in prior work by dissolving amorphous $As_2S_3$ powder in propylamine [11]. This solution is then spin-casted or drop coated on a substrate (TEM grid or silica microscope slide) and thermally processed at different temperatures as specified in the results. Fig. 1a shows the optical image of $As_2S_3$ drop casted on a TEM grid (carbon coated, copper grid). The TEM grid size is 10 µm by 10 µm. A 532nm laser beam and Gaussian beam profile is concentrated in the middle of the film, and the bright circles in Fig. 1a indicate the extent of the laser beam. The dark spot in the center of the laser exposed region shows the laser crystalized area. TEM (Phillips CM200) is performed at 200 keV in bright-field, dark-field, and diffraction conditions. The diffraction pattern of unexposed material shows blurry rings indicating the amorphous nature of the solution processed $As_2S_3$ (Fig. 1b). In comparison, the diffraction pattern from the center of laser-exposed area shows clear crystalline structure formed under both low (40µW/µm2) (Fig. 1c) and high (160µW/µm2) (Figure 1d) laser intensity. The measured lattice spacings for the low intensity exposure (2.636 Å, 1.512 Å) correspond to polycrystalline $As_4S_4$ (2.636 Å, 1.513 Å). Also, the polycrystalline structure shows a preferred orientation indicated with an intensity enhancement in the diffraction as shown in Fig. 1c (marked with a circle). The TEM diffraction pattern of high laser intensity exposure shows distinct periodic spots indicating a single crystal type structures (Fig. 1d). As a small aperture is placed covering only the diffracted beams in the circled region in Fig. 1c, a dark-field image (Fig. 1e) is formed which shows a Morie pattern of ~10 nm spacing due to the overlap of $As_4S_4$ nanostructure crystal layers with preferred orientations.

Bruker D8 Advance X-ray diffractometer was used to obtain pair distribution functions (PDFs) of the $As_2S_3$ system in order to probe its atom-atom correlations. The $As_2S_3$ solution is dried in a ceramic pestle and resulting powder is loaded into a 2.5-inch long polyimide capillary and sealed using quick-setting transparent epoxy. The material is then uniformly exposed to a 30µW/µm$^2$ (low intensity) 532 nm laser for approximately 5 minutes and 120µW/µm$^2$ (high intensity) for the same duration. Ag Kα radiation is used, which has a smaller wavelength compared to Cu radiation, and hence provides a broader $Q$ range. A step size of 0.050°, along with a count time per step of 30s, and a 2$\theta$ range of 3° to 130° is used. The PDF, G(r), was obtained by taking a sine Fourier transform of the measured total scattering function, S(Q), as shown in equation (1) [18]

$$G(r) = \frac{2}{\pi} \int_{Q_{min}}^{Q_{max}} Q[S(Q) - 1]\sin(Qr)dQ \qquad (1)$$

where $Q$ is the momentum transfer given by $Q = 4\pi sin\theta/\lambda$. Standard data reduction procedures are followed to obtain the PDF using PDFgetX2 [19], with a $Q_{max}$ of ~14.4 Å$^{-1}$. Prior to calculation of the PDF, the total scattering function is multiplied by a Lorch window function [20] to improve the signal/noise at the expense of real-space resolution.

Fig. 2 displays the X-ray PDF data for the solution processed sample (SP) together with the laser exposed samples at two different intensities. In contrast to the TEM results, where it was clear that laser exposure led to local alterations of the atomic structure (Fig. 1), the PDF data shows minimal changes to the amorphous atomic structure as a result of laser exposure. The first atom-atom correlation in the X-ray PDF at 2.26 Å is assigned to the As-S bond length (denoted as (i) in Fig.

2), while the second correlation at 3.45 Å is attributed to the atom-atom distances of As-As, S-S, and 2nd nearest-neighbor As-S (denoted as (ii) in Fig. 2) [21]. Slight changes to the ratio of the peak intensities at these locations, *G(i) / G(ii)*, as a result of laser exposure, is indicative of rearrangements to the atomic structure. As the intensity of laser exposure increases, the *G(i) / G(ii)* ratio increases, implying that some of the S-S homopolar bonds are being converted to As-S heteropolar bonds. This observation agrees with the electron diffraction patterns shown in Fig. 1, where upon laser exposure the structure becomes more ordered with a layered anisotropic structure. Hence, the PDF results indicate that exposure to radiation causes the structure to rearrange, although no sign of crystallization is visible in the data using this bulk analysis technique.

Fig. 3a shows Fourier transform infrared spectra (FTIR) of as-spin coated arsenic sulfide without thermal or laser processing, and the same material followed by thermal annealing at 110 °C, 160 °C, and 185 °C, respectively. The broad absorption at 2200-3100 cm$^{-1}$, indicative of amine molecules, gradually weakens as the material is annealed at higher temperature. The two absorption peaks in the range of 1300 – 1700 cm$^{-1}$, representing solvent species in the sample, disappear entirely at the annealing temperature of 185°C. The Differential Scanning Calorimetry (DSC) thermograms of solution-processed $As_2S_3$ powders (red curve in Fig. 3d) exhibits a broad endothermic peak around 135 °C, while the same sample with pre-annealing at 120 °C for 5 hours does not (blue line in Fig. 3d). We expect that the endothermic peak corresponds to the decomposition of sulfur bonds associated with the propylamine dissolution and the removal of H in the form of $H_2S$ as proposed by Chern et al [22]. In addition, we find that temperature around 180 ~ 190°C induced another structural change of the material, as supported by noisy peaks in DSC trace of solution-processed $As_2S_3$ (Fig. 3d). In comparison to the DSC trace of raw $As_2S_3$ (black curve in Figure 3d), this temperature matches the expected glass transition ($T_g$) (Fig. 3e) at which the material becomes liquid like from the glassy state. $T_g$ in raw material is measured to be near 191.7 °C. The structural changes captured by DSC are consistent with the FTIR measurement results.

The PL spectra were collected by coupling the light scattered from the sample to a Horiba Raman spectrometer through a 100× objective, focusing the probe laser (532 nm) spot to ~1.5 µm². The excitation laser intensity is controlled below 0.4µW/µm² to minimize nonlinear response and minimize any subsequent modifications to the material. The presence of solvent in $As_2S_3$ prevents photoluminescence and only samples annealed above 140°C exhibit a strong photoluminescence signal. As the glass restructures with thermal annealing at 160°C and 185°C, the defects levels in solution processed $As_2S_3$ give rise to a broadband emission spectrum in the visible band, centered at 680nm. In the band diagram inset of Fig. 3b and 3c, the black arrow conceptualizes the $As_2S_3$ absorption with near bandgap energy, and the grey arrows show photoluminescence from the defects levels. Laser crystallized $As_4S_4$ has particular defect energy levels compared to thermally processed samples and the PL spectrum shows significant differences (Fig. 3c vs 3b). In particular, the laser irradiated samples show a red shift in the center of the PL spectrum from 680 nm for the thermally processed samples to 750 for the laser processed samples.

Thermal annealing removes solvent species to form isotropic amorphous films, while the polarized laser crystallizes the chalcogenide material (Fig. 1e). The anisotropy of the crystal would be expected to exhibit an excitation polarization dependence in the PL spectrum. After crystallization of the material by high intensity laser (532nm), we probe the modified PL of the crystallized region on the same set-up configuration, but reduce the laser power to monitor the PL spectrum. Figure 4a shows the PL spectrum from laser crystallized materials under different excitation polarizations of 0º, 45º, 90º, 135º and 180º. The polarization degree is relative to the one of crystallization laser. We see that the maximum emission intensity occurs for 90º showing a 50% enhancement of photoluminescence relative to the parallel polarization (0º) further confirming the preferred anisotropy in the polycrystalline structure shown in Fig. 1c.

We studied the localized atomic structure change in solution processed arsenic sulfide by near bandgap laser exposure. Fundamental differences between thermal and laser-based processing are observed by thermal, optical, X-ray and electron microscopy characterization corresponding to differences in the structural properties of the material. These locally crystallized anisotropic chalcogenide materials could be incorporated for designing new components in integrated photonic devices.

AUTHOR INFORMATION

**Correspondent Author**

*E-mail: cbarnold@princeton.edu

**Notes**

The authors declare no competing financial interest.

**ACKNOWLEDGMENT**

The authors thank intriguing discussions with Prof. Yueh-Lin Loo of Princeton University, and facility at Imaging and Analysis Center at Princeton University. This work was supported by NSF through the MIRTHE Center (Grant No. EEC-0540832) and MRSEC Center (Grant No. DMR-1420541).

**Figures:**

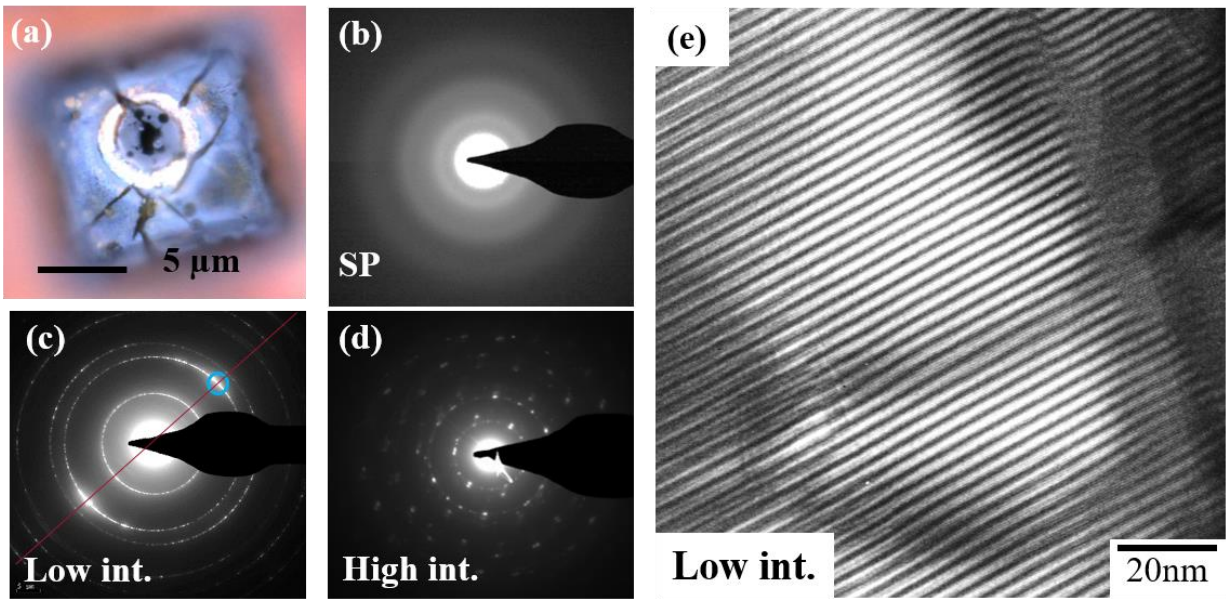

Figure 1. Laser Gaussian beam induced pattern on chalcogenide thin film. (a) Optical image of solution processed As$_2$S$_3$ film drop-casted on TEM grid, observed under 100x microscope after laser exposure of 10 second, with average power of 160µW/µm$^2$. Scale bar: 5µm. (b) Electron diffraction pattern for the unexposed region and (c) Crystallized region near the center of the beam (40µW/µm$^2$). The red line shows the preferred nanocrystal orientation. (d) The crystallized region after laser exposure with 160µW/µm$^2$. (e) A corresponding dark-field image showing the real-space structure as in (c), through the aperture in the blue circle.

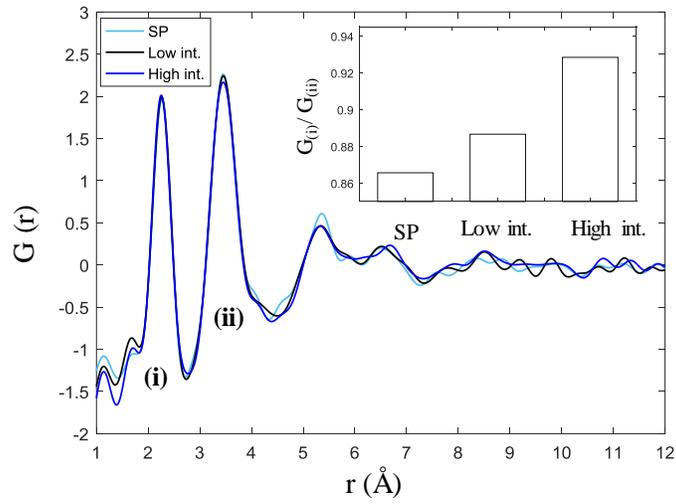

Figure 2. Laboratory X-ray PDF data for solution processed sample (SP), after low intensity laser exposure (532nm, 30µW/µm$^2$) and high intensity laser exposure (532nm, 120µW/µm$^2$), Inset: The peak intensity ratio of the first nearest neighbor correlation intensity (i) to the second (ii).

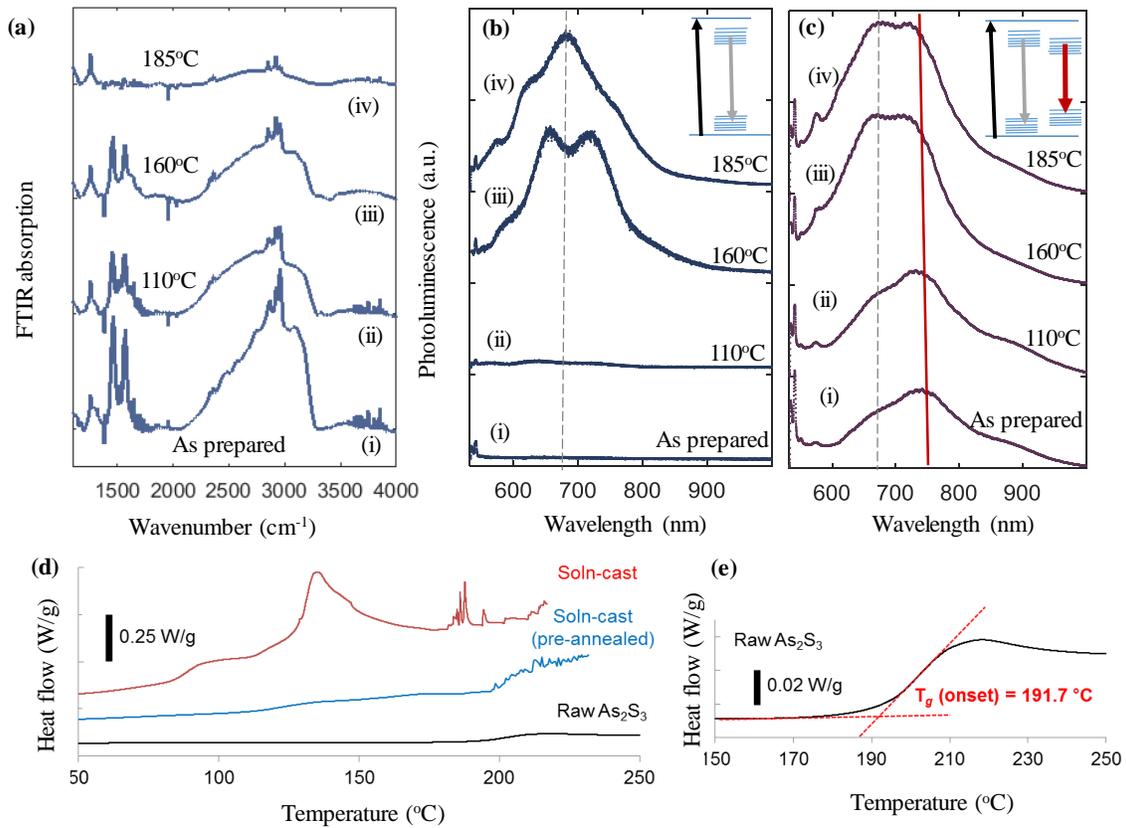

Figure 3. Photoluminescence of laser and thermal induced defects. (a) FTIR measurement of spin coated sample before (i) and after annealing in nitrogen environment at (ii) 110°C (iii) 160°C and (v)185°C for 5 hours. Note, curves all have same scale but are shifted on the y-axis for clarity. (b) Photoluminescence of corresponding samples shown in (a). Grey dashed line marks the center wavelength of the thermal induced defects state. Inset: band diagram and absorption (black arrow) and emission (grey arrow) from the thermal induced defects. Spectra are individually normalized and shifted for clarity. (c) Photoluminescence of spin coated sample after 40μW/μm$^2$ laser exposure for 10s. The grey dashed line and red solid line mark the center wavelength of amorphous thermally annealed sample and laser crystallized structure, respectively. Inset: absorption (black arrow), emission from thermal induced (grey arrow) and laser induced (red) defects. (d) Differential scanning calorimetry (DSC) thermograms of raw (black), solution processed (blue)

and pre-annealed (120°C for 5 hours in vacuum of solution processed sample) arsenic sulfide. The heating rate is 10°C/min. (e) Magnified DSC measurement showing the glass transition of raw arsenic sulfide.

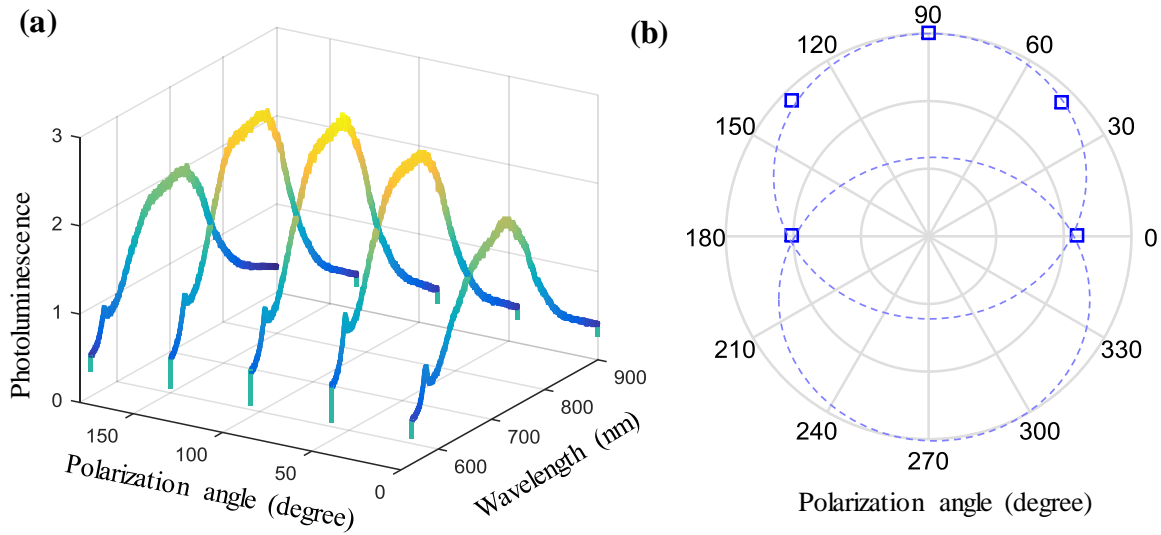

Figure 4. Photoluminescence spectra of the laser crystallized chalcogenide. (a) Photoluminescence spectra, (b) intensity versus relative polarization angle to the crystallization laser. The blue squares are experimental data and the dashed circles are fitted circles.